\documentclass[aps,pre,showpacs,twocolumn,superscriptaddress]{revtex4}
\usepackage{graphicx,bm,amssymb}
\begin{document}

\title{Theory of the isotropic-nematic transition in dispersions of compressible rods}
\author{Kostya Shundyak}
\affiliation{Instituut-Lorentz for Theoretical Physics, Leiden University, Niels Bohrweg 2,
2333 CA Leiden, The Netherlands}
\author{Ren\'e van Roij}
\affiliation{Institute for Theoretical Physics, Utrecht University, Leuvenlaan 4, 3584 CE
Utrecht, The Netherlands}
\author{Paul van der Schoot}
\affiliation{Eindhoven Polymer Laboratories, Technische Universiteit Eindhoven,
P.O. Box 513, 5600 MB Eindhoven, The Netherlands}
\date{February 2, 2006}

\begin{abstract}
We theoretically study the nematic ordering transition of rods that are able
to elastically adjust their mutually excluded volumes. The model rods, which
consist of a hard core surrounded by a deformable shell, mimic the structure
of polymer-coated, rod-like fd virus particles that have recently been the
object of experimental study [K.~Purdy et al., Phys. Rev. Lett. \textbf{94}, 
057801 (2005)]. We find that fluids of such
soft rods exhibit an isotropic-nematic phase transition at a density higher
than that of the corresponding hard-rod system of identical diameter,
and that at coexistence the order parameter of the nematic phase depends
non monotonically on the elastic properties of the polymer coating.
For binary mixtures of hard and soft rods, the
topology of the phase diagram turns out to depend sensitively
on the elasticity of shell. The lower nematic-nematic critical point,
discovered in mixtures of bare and polymer-coated fd virus particles, is not
reproduced by the theory.
\end{abstract}

\maketitle

\section{Introduction}
Recently developed methods to control the contour length and the
effective diameter of elongated virus particles
\cite{DFPTRSLA01,PVGJFPRL05} have opened up the way to
systematically study the bulk phase behavior of mono- and
bidisperse rods over a relatively wide range of lengths and
widths. This is important, because it allows for the experimental
verification of a vast amount of theoretical work that has been
done on the isotropic-to-nematic (\textit{IN}) phase transition in
dispersions of mutually repelling rods, and in particular that on
bidisperse mixtures \cite{AFM78,OM86,LCHDJCP84,BKPVSSA88,VLRPP92,SSJCP02}.
The agreement of theoretical predictions for binary dispersions with experimental
data \cite{ITM84,ITPJ84,SIITP89,STAP94,BKLJPCM96,KLPRL00}
has so far not been as impressive as that for monodisperse ones
\cite{O,STAPS96}, where the impact of, e.g., the size~\cite{FJPC87}, shape~\cite{BFJCP97},
molecular flexibility~\cite{ITPJ88,ITM88,DSJCP91} and Coulomb
interactions~\cite{STPA91,PLKKPRE02,POKRPRE05}
appear to be well understood \cite{VLRPP92,Fbook}.

One of the reasons for this state of affairs is probably that in
binary mixtures there is a coupling between ordering,
fractionation, and demixing, leading to a much more complex phase
behavior \cite{VLRPP92,SJJCP95,RMJCP96,RMDP98,SSJCP02}.
Indeed, in bidisperse systems more length scales
compete with each other, presumably making them more sensitive to
the effects of flexibility, residual Van der Waals attractions,
non-additivity or charges \cite{MFM81,OM86,HSJCP92,SOJCP92,SRSJCP05}.
Binary mixtures are therefore a
much more critical gauge of the accuracy of theories than
monodisperse systems are. In some cases, the disagreement between
theory and experiment is not just quantitative~\cite{STAPS96} but
even qualitative. For example, in experiments on aqueous mixtures
of bare fd virus particles and fd virus particles onto which
polymeric chains are grafted, Purdy et al. \cite{PVGJFPRL05}
discovered a nematic-nematic coexistence region that exhibits a
lower critical point. Extensions of Onsager's classical
second-virial theory for infinitely rigid rods incorporating a
diameter bidispersity do not predict such a lower critical point
\cite{LCHDJCP84,RMJCP96,RMDP98,SRPRE03}, not even if one allows
for non-additivity of the interactions between the species
\cite{SRSJCP05}. However, these theories do predict a
nematic-nematic demixing either with an upper critical point or no
critical endpoint at all \cite{VLJPC93,RMEL96,RMDP98,SSJCP02}.

Theoretically, nematic-nematic demixing transitions of bidisperse
rods with a lower critical endpoint have been predicted, but only
for rods with a sufficiently large bending flexibility
\cite{SSVSA89}, or for sufficiently short rods for which higher
order virials become important \cite{SMCLC73,SOJCP92,WVLJPCB01,VGJMP03,VPGFJPRE05,Rprep}.
It appears, however, that some of the predictions are quite sensitive to the invoked
approximations, and in our view the issue remains contentious. For
instance, depending on how precisely higher order virials are
approximately accounted for, the lower critical point appears and
disappears in the phase diagram \cite{SRprep}. This is
reminiscent of the qualitative difference in the predicted phase
behavior of hard-rod mixtures, depending on whether the exact
orientational distribution is used or a Gaussian approximation to
that \cite{VLJPC93,HMP99,RMEL96,RMDP98,SSJCP02}. So, given the
apparently inherent sensitivity to the details of the theories, it
remains important to explore alternative explanations of the
observed phase behavior, and in particular that of the lower
critical end point.

A factor that has not been considered, and that could
significantly influence the phase behavior in systems such as
those studied by Purdy and coworkers, is the finite
compressibility of the polymer coating of the rods. The effective
diameter of these polymer-coated rods is not necessarily fixed but
could well be a function of the thermodynamic state of the
dispersion. So far, the dimensions of rod-like colloids have been
treated as quenched variables, i.e., as invariants of the
thermodynamic state of the suspension. (An exception to this are
micellar rods with annealed length distribution \cite{SCL94}.)
With the polymer-coated rods of Ref.~\cite{PVGJFPRL05} in mind, it
seems opportune to explicitly consider the elastic response of
this coating to the osmotic pressure of the dispersion. Contrary
to for instance theoretical work on the crystallization of
dendrimers, where entropic interactions of a similar nature are
modelled by a soft potential \cite{JWLM99}, we model the effects
of compression not at the interaction potential level but at the
level of the volume exclusion. In effect, our theory is that of
particles with an annealed diameter, and hence an annealed
excluded volume.

Our calculations show that monodisperse fluids of elastically
compressible rods exhibit an isotropic-nematic phase transition at
a density higher than that of the corresponding incompressible-rod
system of equal diameter, as expected. How much higher,
depends on the stiffness of the shell. Weakly impacted
upon by the shell stiffness is the relative density gap
in the coexisting phases. For binary mixtures of incompressible
and compressible rods, the structure of
the phase diagram changes dramatically with varying elasticity of
the coating, showing once more the sensitivity of this kind of
mixtures to the details of the interactions \cite{SRSJCP05}.
Still, the topology of the phase diagrams that we calculate does
bear some resemblance to that of incompressible hard-rod mixtures
of unequal diameter: an upper but no lower nematic-nematic
critical point is produced by the theory.

The remainder of this paper is organized as follows. In
Sec.~\ref{sectfunct} we first introduce the Onsager-type free
energy functional, and derive from that the basic Euler-Lagrange
equations describing the orientational and density distribution of
the elastically compressible rods under conditions of
thermodynamic equilibrium. In Sec.~\ref{monosect} we study the
behavior of a monodisperse fluid of such compressible rods. In
order to obtain an analytical solution to the model we invoke a
Gaussian approximation to the orientational distribution function.
This analytical theory we compare with an exact, numerical
evaluation and find fair agreement. In Sec.~\ref{binsect}, we
analyse the bulk phase diagrams of binary mixtures of hard rods
and elastically compressible ones, but now only numerically
keeping in mind the sensitivity of the binary phase diagram to
approximations invoked. A summary and discussion of the results
are presented in Sec.~\ref{nasummarysect}.

\section{Density functional and method}\label{sectfunct}
We are concerned with the bulk properties of a
fluid of cylinders of two different species $\sigma=1,2$ of
diameter $D_{\sigma}$ and equal length $L$ in a macroscopic volume
$V$ at temperature $T$ and chemical potentials $\mu_{\sigma}$. We
presume the limit $D_{\sigma}/L\rightarrow0$ to hold, in which
case a second virial theory is believed to be exact \cite{comIN}. The
``effective'' diameter $D_{1}$ of the thin rod is determined by the
bare hard core of the particle $D_{1}=\Delta^{core}$, whereas the
diameter of the thick (coated) rods is written as
$D_{2}(\gamma)=\Delta^{core}+\gamma\Delta^{pol}$ with
$\Delta^{pol}$ twice the thickness of the soft polymeric shell.
The compression factor $\gamma\in\lbrack0,1]$ parameterizes the
deformation of this soft shell due to interactions with
other rods in the system. At infinite dilution we expect
zero deformation of the thick rigid rod, i.e., $\gamma=1$,
and it is convenient to define the limiting diameter ratio
$d=D_2(\gamma=1)/D_1=1+\Delta^{pol}/\Delta^{core}$.

The compression of the grafted polymer layer reduces the excluded volume,
as we will see explicitly below, at the expense of an elastic energy.
Introducing the rigidity $k$, we write the elastic energy of a
single rod at a compression equal to $\gamma$ as
$k(\gamma^{2}+\gamma^{-2}-2)/2$ in units of thermal energy
$k_BT$. It is inspired by the theory of ideal (Gaussian)
polymers \cite{GKbook}, where we presume that the grafted
polymers are not strongly stretched \cite{PVGJFPRL05}.
Hence, we expect the numerical value of the
rigidity $k$ to be of order of the number of chains grafted onto
each rod. However, for simplicity we presume it
to be a free parameter. Note that for $\gamma=1$ the elastic
contribution to the free energy is zero, as it should.

The total grand potential $\Omega[\rho_1,\rho_2]$ of the
spatially homogeneous suspension is now written as a functional of the distribution
functions $\rho_{\sigma}({\bf u})$, where ${\bf u}$ denotes the
unit vector along the axis of a rod.
The distributions are normalized such that $\int d{\bf
u}\rho_{\sigma}({\bf u})=n_{\sigma}$, the number density of
species $\sigma$ at the imposed chemical potential. Within the
second virial approximation we write the functional as \cite{O,VLRPP92}
\begin{eqnarray}\label{grandpot}
\frac{\beta\Omega[\{\rho_{\sigma}\}]}{V}&=&
\sum_{\sigma}\int d{\bf u}\rho_{\sigma }({\bf u})
\Big(\ln[\rho_{\sigma}({\bf u})\nu _{\sigma}]-1-\beta\mu_{\sigma}\Big)\nonumber\\
&&+\frac{1}{2}\sum_{\sigma\sigma'}\int d{\bf u}d{\bf u}'
E_{\sigma\sigma'}({\bf u};{\bf u}')\rho_{\sigma}({\bf u})\rho_{\sigma'}({\bf u}')\nonumber\\
&& +\frac{1}{2}k\left(\gamma^{2}+\gamma^{-2}-2\right)\int d{\bf u}\rho_{2}({\bf u}),
\end{eqnarray}
with $\beta=(k_{B}T)^{-1}$ the inverse temperature, $\nu_\sigma$ the thermal
volume of the species $\sigma$, and the
excluded volume due to hard-core interactions
\begin{eqnarray}\label{exclvolnonadd}
E_{\sigma\sigma'}({\bf u},{\bf u}')=L^{2}(D_{\sigma}+D_{\sigma'})
|\sin (\arccos({\bf u}\cdot{\bf u}'))|,
\end{eqnarray}
where additional $O(LD^{2})$ terms are being ignored,
in line with the needle limit ($D_{\sigma}/L\rightarrow0$) of interest here.

The equilibrium conditions on the functional,
$\delta\Omega[\{\rho_{\sigma}\}]/\delta \rho_{\sigma}({\bf
u})=0$ and $\partial\Omega[\{\rho_{\sigma}\}]/\partial
\gamma=0$, lead to the set of nonlinear integral equations,
\begin{eqnarray}\label{nonlinset}
\beta\mu_{\sigma}&=&\ln[\rho_{\sigma}({\bf u})\nu_{\sigma}]+
\sum_{\sigma'}\int d{\bf u}'E_{\sigma{\sigma
}^{\prime}}({\bf u},{\bf u}^{\prime})\rho_{{\sigma}^{\prime}}({\bf u}^{\prime})\nonumber\\
&+&\frac{1}{2}\delta_{\sigma,2}k\left(  \gamma^{2}+\gamma^{-2}-2\right),\\
0&=&k\left( \gamma-\gamma^{-3}\right)\int d{\bf u}\rho_{2}({\bf u})\nonumber\\
&+&\frac{1}{2}\sum_{\sigma{\sigma}^{\prime}}\int  d{\bf u}d{\bf u}^{\prime}
\frac{\partial E_{\sigma{\sigma}^{\prime}}({\bf u},{\bf u}^{\prime})}{\partial\gamma}\rho_{\sigma}({\bf u}
)\rho_{{\sigma}^{\prime}}({\bf u}^{\prime}),\nonumber
\end{eqnarray}
to be satisfied by the equilibrium distributions. The
$\gamma$-derivative in the last line involves the
$\gamma$-dependence of $D_2$ through Eq.~(\ref{exclvolnonadd}).
These equations can be solved, either approximately within a
Gaussian approximation or numerically on a grid of orientations.
Details of the numerical schemes have been discussed
elsewhere \cite{RMDP98,SRPRE03}. Here, in order to find the bulk
uniaxially symmetric distributions $\rho_{\sigma}(\theta_{i})$,
with $\theta=\arccos({\bf u}\cdot{\bf n})$ the angle
between the rod unit vector ${\bf u}$ and the nematic
director ${\bf n}$, we use a nonequidistant
$\theta$-grid of $N_{\theta}=30$ points $\theta_{i}\in [0,\pi/2]$,
$1\leq i\leq N_{\theta}$, with $2/3$ of them uniformly distributed in
$[0,\pi/4]$.
The resulting distributions can be inserted into the functional to
obtain the grand potential $-pV$, with $p$ the pressure. Then the
complete thermodynamics can be inferred, as well as the phase
diagram.

\section{Monodisperse soft rods}\label{monosect}
As demonstrated in Refs.~\cite{O,VLRPP92}, the $IN$ transition in
suspensions of rigid rods is a result of a competition between the
orientational entropy and the packing entropy (free volume).
In order to estimate the impact of the elastic term of Eq.~(\ref{grandpot})
on the $IN$ transition, we first restrict
our attention in this section to a purely monodisperse system of coated rods.
Formally, this can be achieved by considering the limit
$\beta\mu_1\rightarrow-\infty$ and $\rho_1({\bf
u})L^2D_1\rightarrow 0$ in the functional and its minimum
conditions, and in this paragraph we drop the species index ``2'' for
convenience.

First, instead of numerically solving for the minimum
conditions we adopt a Gaussian Ansatz for the one-particle
distribution function in the nematic phase, with $\rho({\bf
u})\propto\exp(-\alpha\theta^{2}/2)$ for $\theta \leq \pi /2$,
and $\rho({\bf u})\propto\exp(-\alpha (\pi-\theta)^{2}/2)$
for $\theta \geq \pi /2$. Here, $\alpha$ denotes a variational
parameter that we fix by minimizing the free energy. From Eqs.~(\ref{nonlinset})
we obtain the following relations between the density and the
compression in the isotropic phase $I$, where $\alpha\equiv0$,
and in the nematic phase $N$, where $\alpha=4c_N^2/\pi$ in terms
of the dimensionless concentration $c_N$ defined below:
\begin{eqnarray}\label{gauss}
\beta\mu_I&=&\ln c_I+2c_I\frac{D(\gamma_I)}{D(1)}
+\frac{1}{2}k\left(\gamma_I^{-2}+\gamma_I^{2}-2\right),\nonumber\\
c_I&=&\frac{kD(1)}{\Delta^{pol}}\left(\gamma_I^{-3}-\gamma_I\right),\\
\beta\mu_{N}&=&3\ln c_N+2\ln\left(\frac{D(\gamma_N)}{D(1)}\right)
+\frac{k}{2}\left(\gamma_N^{-2}+\gamma_N^{2}-2\right)+C,\nonumber\\
2&=&\frac{kD(\gamma_N)}{\Delta^{pol}}\left(\gamma_N^{-3}-\gamma_N\right),\nonumber
\end{eqnarray}
with $C=2\ln2\pi^{-1/2}+3$,
$c_{I(N)}=(\pi/4)n_{I(N)}L^{2}D(1)$ the
dimensionless number density of the $I$ ($N$) phase, $D(\gamma
)=\Delta^{core}+\gamma\Delta^{pol}$ the effective diameter of the
coated rods, and $D(1)=\Delta^{core}+\Delta^{pol}$.
In addition, the dimensionless pressures
$p^*=(\pi/4)\beta pL^2D(1)$ of the isotropic
and nematic phases can be written as
\begin{eqnarray}\label{gaussp}
p^*_I=c_I\left(1+c_I\frac{D(\gamma)}{D(1)}\right),&&
p^*_N=3c_N.
\end{eqnarray}

\begin{figure}[t]
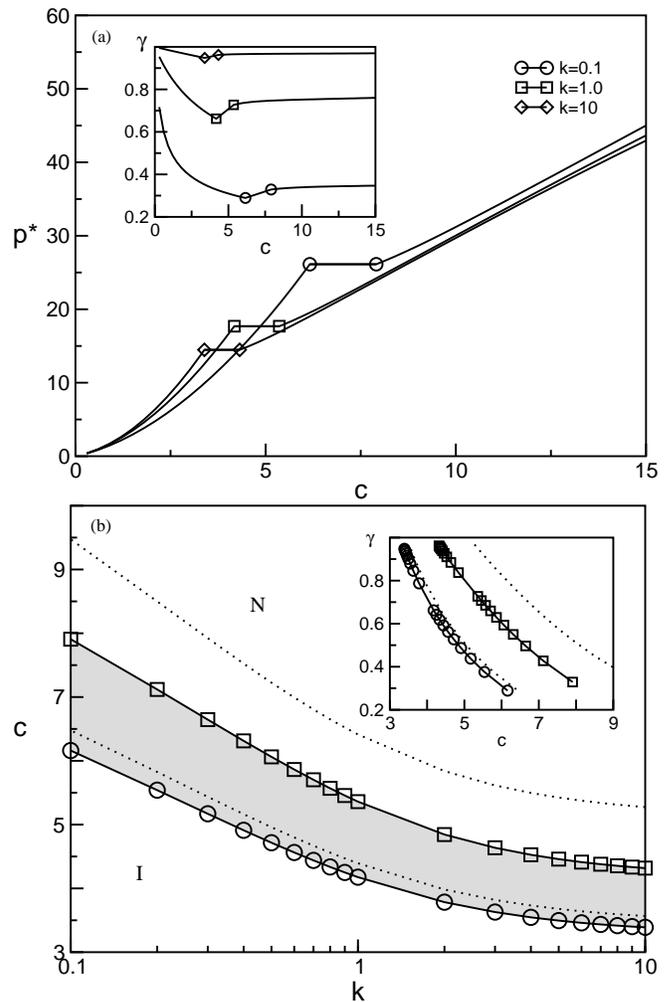

\centering
\includegraphics*[width=8.6cm]{eqstate.eps}
\includegraphics*[width=8.6cm]{pcD3.eps}
\caption{\label{phaseD3} \small
(a) Equation of state for a fluid of monodisperse soft rods
($d=1+\Delta^{pol}/\Delta^{core}=3$) in terms of the dimensionless
bulk pressure $p^*=(\pi/4)\beta pL^2D(1)$ and the dimensionless density
$c=(\pi/4)nL^{2}D(1)$ for several values of shell rigidities $k=0.1,1.0,10$.
The inset shows the dependence of the compression factor $\gamma$
on density $c$ for the same values of $k$ as in the main plot.
(b) Bulk phase diagram for the same system in $c-k$ coordinates.
The coexistence region (grey area) separates regions of the isotropic ($I$) and
nematic ($N$) phase, and the tie-lines connecting coexisting phases are
vertical. The dotted lines indicate the corresponding coexistence $IN$ curve,
calculated with the Gaussian trial function (Eqs.~(\ref{gauss}),~(\ref{gaussp})).
The inset shows the dependence of the compression factors $\gamma_{I,N}$
on density $c$ at $IN$ coexistence for the same range of $k$ as in the main
plot.}
\end{figure}
Note that in the isotropic phase (i) the compression $\gamma_{I}$
decreases monotonically with increasing $c_{I}$, and (ii) the
elastic contribution to the free energy grows with increasing
density, as one might in fact expect. More interestingly, in the
nematic phase the rigidity and geometric parameters of the rods
fully determine their compression, which is independent of
concentration (at the level of the Gaussian approximation). Hence,
in the nematic phase the contribution of the elastic compression
can be considered as an effect of an uniform bulk field that
renormalizes the value of the chemical potential. This is also
consistent with expression for the pressure of the nematic phase
which is only indirectly affected by the elasticity of the polymer
coating of the rods. Overall, the elastic term shifts
the \textit{IN} transition to higher
densities due to reduction of the effective diameter $D(\gamma)$.
The results of our calculations
are shown in Fig.~\ref{phaseD3} for the system
with $d=3$, and other values of $d$ lead to the similar phase diagrams.

In Fig.~\ref{phaseD3}(a) we show dependence of the dimensionless bulk pressure
$p^*$ on the dimensionless number density $c$ (i.e., the equation of state) for several values
of rigidity $k$ of the polymeric shell of the rod. The solid lines represent
direct numerical solutions of Eqs.~(\ref{nonlinset}).
The results of calculations within the Gaussian approximation are
quite close to those from our numerical calculations, albeit that they overestimate
the rod densities in the $I$, $N$ phases at coexistence
(not shown here for the sake of clarity).

\begin{figure}[t]\centering
\includegraphics*[width=8.6cm]{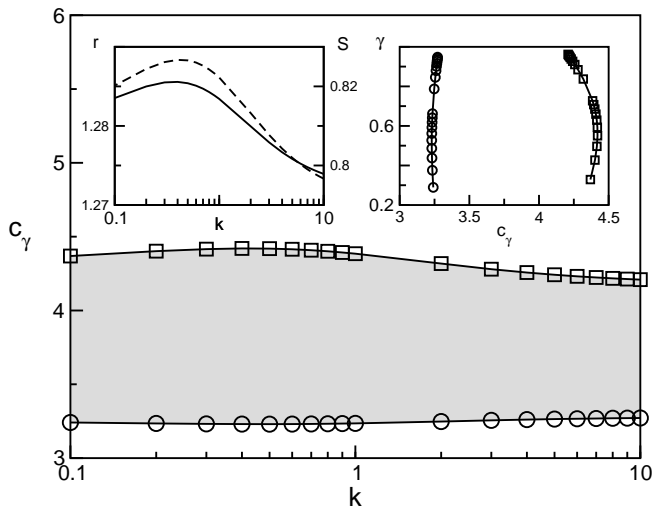}
\caption{\label{phaseD3scaled} \small
The same bulk phase diagram as in Fig.~\ref{phaseD3}(b)
in terms of the scaled density $c_\gamma =(\pi/4)nL^{2}D(\gamma)$
versus rigidity of the polymeric shell $k$.
The right inset shows the dependence of the compression factor $\gamma$
on the density $c_\gamma$ for the same values of $k$ as in the main
plot. The left inset shows the ratio $r=c_{\gamma,N}/c_{\gamma,I}$
of the densities of coexisting $N$ and $I$ phases (solid line)
and the nematic order parameter $S$ (dashed line) as a function of $k$.
The right inset gives the compression $\gamma$ of the rods
in the coexisting phases, as a function of $c_\gamma$.}
\end{figure}
Imposing conditions of mechanical and chemical equilibrium between
isotropic and nematic phase, we calculated the densities and
the compression factors of the rods in the $I$ and
$N$ phases at coexistence, and present these in Fig.~\ref{phaseD3}(b)
for the realistic range of rigidities of $k\in[0.1,10]$.
The dotted lines give the results of the calculations within the Gaussian approximation,
whereas the numerical solutions to Eqs.~(\ref{nonlinset})
are represented by the symbols, with the solid lines serving as a guide to the eye.
Whilst the predictions for the concentration of rods in the isotropic phase
at coexistence are in almost quantitative agreement,
those for the phase gap are not: the Gaussian approximation overestimates
the phase gap by a factor of about two. For rigidities of order $k\approx10$ the limiting densities
$c_{IN}$ of the co-existing phases approach the values of the corresponding hard rod system.
The polymeric shells of the rods are then only slightly deformed ($\gamma\approx 1$).
On the other hand, in the limit of small $k$, the densities at coexistence approach
those of hard rods with the smaller hard-core diameter $\Delta^{core}$,
because in that case $\gamma\rightarrow 0$. The leveling off occurs only for very small
values of $k$, and is not shown in  Fig.~\ref{phaseD3}(b).
The compression factors $\gamma_{I,N}$ of the rods in the coexisting phases
are shown in the inset. The values calculated within the Gaussian approximation
are similar to those obtained from our numerical analysis.
Again, in the isotropic phase the agreement is almost quantitative.

\begin{figure}[!t]\centering
\includegraphics*[width=8.4cm]{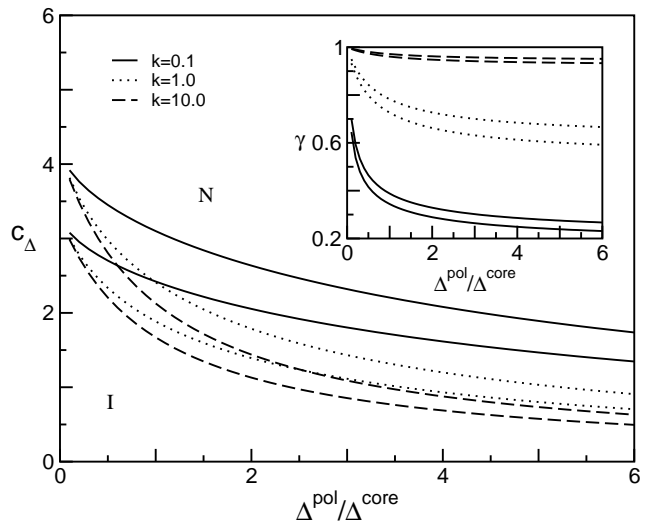}
\caption{\label{cDelta} \small
Bulk phase diagrams in terms of the dimensionless density
$c_\Delta =(\pi/4)nL^{2}\Delta^{pol}$ versus relative
thickness of the polymeric shell $\Delta^{pol}/\Delta^{core}$
for rods with several rigidities $k=0.1,1.0,10$.
The inset shows the dependence of the compression factors $\gamma$
in coexisting phases on $\Delta^{pol}/\Delta^{core}$
for the same values of $k$ as in the main plot.}
\end{figure}
It is clear from Fig.~\ref{phaseD3} that the dimensionless density $c$
at the $IN$ transition increases with a softening of the polymer layer,
which of course is not all that surprising.
From a theoretician's point of view, $c=(\pi/4)nL^{2}D(1)$
is indeed the preferable concentration scale, because it does not
depend explicitly on $\gamma$. The question now arises whether experimentally
the bare diameter $D(1)$ of the rods might be determined from the actual
concentration of particles at which the nematic phase appears \cite{VLRPP92,Fbook}.
The answer appears to be negative, because the actual diameter will not have this
value at the point where the nematic phase becomes stable. If $D(1)$ were determined
independently, e.g., from the compressibility at low concentrations, then this
would allow one to obtain a value for $k$.

\begin{figure*}[!ht]
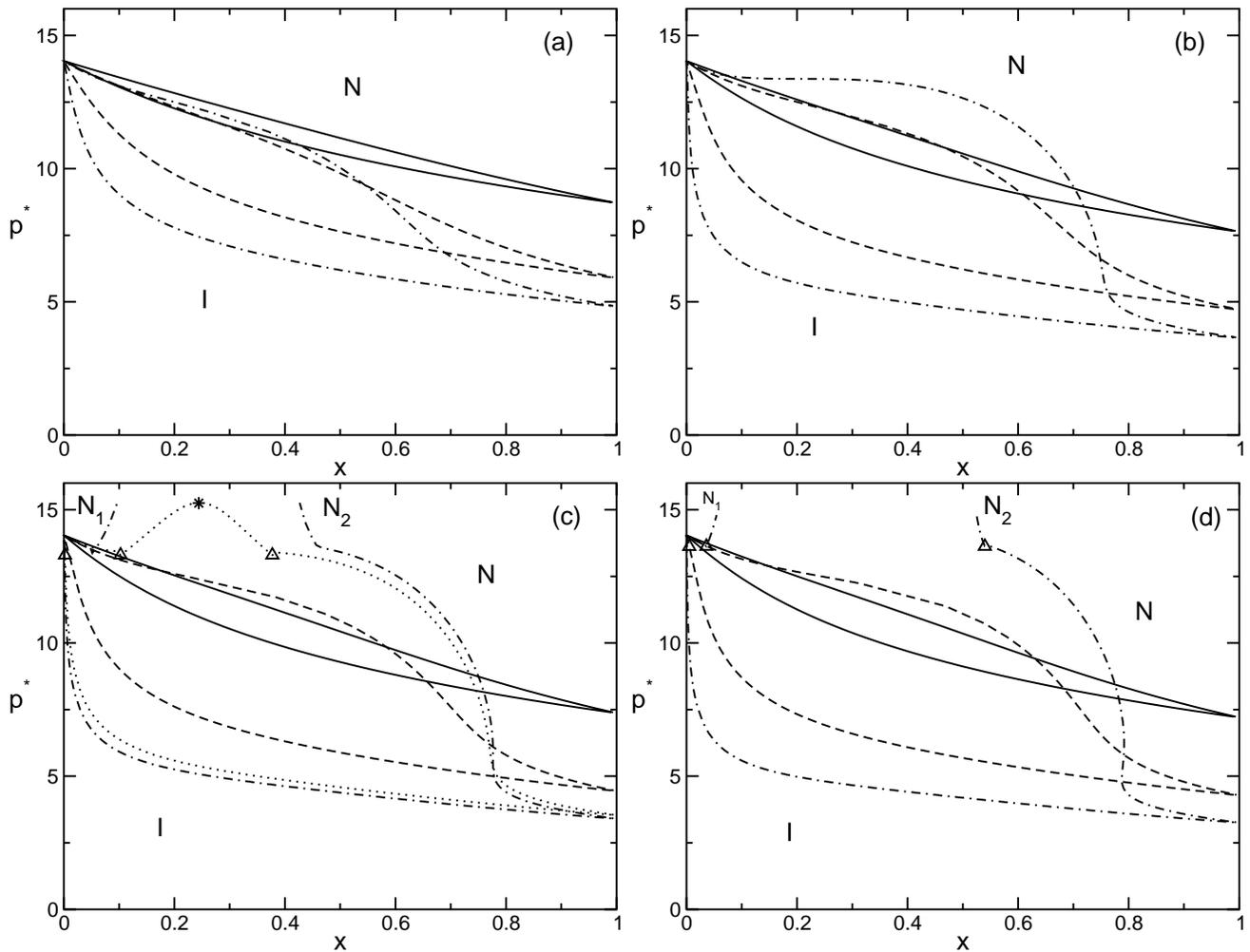
\centering
\includegraphics*[width=8.6cm]{all3.eps}
\includegraphics*[width=8.6cm]{all4.eps}
\includegraphics*[width=8.6cm]{all4_3.eps}
\includegraphics*[width=8.6cm]{all4_5.eps}\caption{{\small Bulk phase
diagrams of binary mixtures of hard ($\Delta^{pol}_{1}=0$) and soft rods
$d=3.0$ (a), $4.0$ (b),$4.3$ (c), $4.5$ (d) in terms of the dimensionless
pressure $p^{*}=(\pi/4) \beta pL^{2}\Delta^{core}$ and the composition $x=n_{2}
/(n_{1}+n_{2})$ for different rigidities: $k=0.1$ (solid), $k=1.0$ (dashed),
$k=5.0$ (dotted, only in (c)), $k=10$ (dot-dashed). In (c,d) $\Delta$ indicate
the isotropic $I$ and nematic $N_{1}$,$N_{2}$ phases of different composition
at coexistence, and the $NN$ remixing point in (c) is marked with $\ast$.}}
\label{bin}
\end{figure*}

Theoretically, the structure of Eqs.~(\ref{gauss}), (\ref{gaussp})
hints at the usefulness of the scaled densities $c_\gamma=(\pi/4)nL^{2}D(\gamma)$,
and it is instructive to compare both density representations
for it helps explain the underlying physics.
The phase diagram of Fig.~\ref{phaseD3} redrawn in terms of $\{c_\gamma,k\}$
is shown in Fig.~\ref{phaseD3scaled}. It suggests that the density
variation with $k$ at the \textit{IN} transition mainly comes from
the reduction of the effective diameter $D(\gamma)$ the rods. This
conclusion is supported by the similarity of the values of the
compression $\gamma_{I,N}$ of the polymer shells of the rods in the
co-existing phases, the dependencies of which on the
density $c_\gamma$ are presented in the right inset in Fig.~\ref{phaseD3scaled}.

Finally, several other quantities can be useful
for qualitative comparisons with experiments.
In the left inset in Fig.~\ref{phaseD3scaled} we present
the density ratio $r=c_{\gamma,N}/c_{\gamma,I}$
of the \textit{I} and \textit{N} phases at
coexistence, as well as the nematic order parameter $S$
as a function of the rigidity $k$. The dependence of
$c_{N}/c_{I}$ of the unscaled densities on $k$ is similar to that
of $c_{\gamma,N}/c_{\gamma,I}$, and is
not shown here. The small variations of $c_{\gamma,N}/c_{\gamma,I}$ and $S$ with
$k$ indicate that in experiments it would be quite hard to use
these two quantities to determine the shell rigidity $k$.

Variation of the molecular weight of polymer that forms the shells of
the rods allows for a modification of its thickness $\Delta^{pol}$.
In Fig.~\ref{cDelta} we show several bulk phase
diagrams in terms of the dimensionless densities
$c_{\Delta}=(\pi/4)nL^{2}\Delta^{core}$ and the ratio $\Delta^{pol}/\Delta^{core}$
for systems with rigidities $k=0.1,1,10$. Note that our previous definition of
the scaled density $c=(\pi/4)nL^{2}D(1)$ (as in Eqs.~(\ref{gauss}),~(\ref{gaussp})
and Fig.~\ref{phaseD3}) would be inconvenient here, as it contains explicit
dependence on $\Delta^{pol}$. At the smallest studied
thickness of the polymeric shell $\Delta^{pol}/\Delta^{core}=0.1$
densities $c_{\Delta,IN}$ closely approach correspondent values for
monodisperse hard rods. As thickness of the shell increases,
the transition densities $c_{\Delta,IN}$ decrease as one would in fact expect.
The inset in Fig.~\ref{cDelta} shows the variation with the shell thickness
of the compression factors $\gamma_{I,N}$ for the same $k$'s as in the main plot.
It appears that the compression of the shells increases for larger values of $\Delta^{pol}$.

\section{Binary mixtures}\label{binsect}
For binary mixtures of soft rods the effective
diameters $D_{\sigma}(\gamma)$ are different, and one can
expects significant modifications of the phase behavior in
comparison with the hard-rod binary fluids. We study the bulk
properties of mixtures of bare hard and coated soft rods with
$d=3.0,4.0,4.3,4.5$ for various values of $k$, chosen to mimic the experiments
of Ref.~\cite{PVGJFPRL05}. Before presenting results of our
calculations, it is useful to recall the structures of the bulk phase diagrams
of hard-rod mixtures, i.e., of mixtures of incompressible rods with different diameters.
For all values of the parameter $d$ they exhibit a low-density $I$ phase,
which at some intermediate values of densities
separates into coexisting $I$ and $N$ phases with
different composition. In mixtures with $d<4.0$ the high-density region has
a single $N$ phase, whereas for the systems with
$d\geq4.0$ this $N$ phase can demix
into two different nematics $N_{1}$,
$N_{2}$ (depending on the composition). The diameter ratio $d$ of the species also determines
whether this $NN$ phase separation persists to (arbitrary) high
densities ($d>4.2$), or whether these $N_1$, $N_2$ phases remix
back~\cite{VLRPP92,VLJPC93,RMEL96}.

The phase diagrams of the ``soft-hard'' mixtures were determined
by solving Eqs.~(\ref{nonlinset}) under conditions of mechanical and chemical
equilibrium. We have verified that an artifact resulting from
the discretization of the angular degrees of freedom of the rods,
and that produces a nematic phase of perfect orientational order,
does not interfere with our calculations \cite{SRPRE04}.

\begin{figure}[t]\centering
\includegraphics[width=8.6cm]{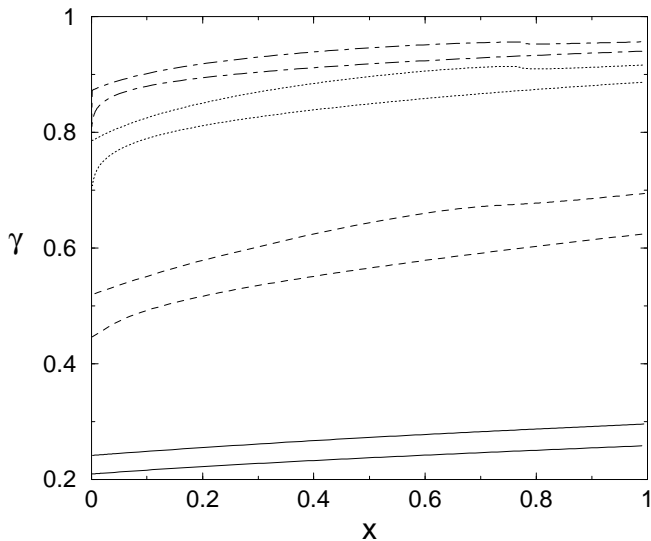}\caption{\small
Compression factors $\gamma$ as a function of the mole fraction $x$
in the coexisting isotropic (lower curve of each pair) and
nematic (upper curve of each pair) phases in the binary mixture
with $d=4.3$. Different pairs of curve correspond from bottom to top
to increasing values of the rigidity $k=0.1,1.0,5.0,10$. See also
Fig.~\ref{bin}(c).}
\label{gam43}
\end{figure}
The results of the calculations are shown in Fig.~\ref{bin},
where we give the bulk phase diagrams of the mixtures
in terms of the dimensionless pressure $p^{*}=(\pi/4)\beta pL^{2}\Delta^{core}$
and composition $x=n_{2}/(n_{1}+n_{2})$ for different
rigidities $k$. In this representation the tie lines connecting
coexisting state points are horizontal because they correspond to
the equal pressure condition. There are several features of the
bulk phase diagrams which we would like to point out:

(i) For all diameter ratios $d$ of mixtures of hard and very soft rods
($k\sim0.1$), the pressure $p^{*}$ at coexistence varies almost
linearly with composition $x_{N}$ of the nematic phase. Detailed
numerical evaluations show that $p^{*}$ depends approximately linearly
on the total density of the mixture $n=n_{1}+n_{2}$, which is
reminiscent to the behavior of a monodisperse system.
This is to be expected, because there is only a weak
coupling then to the elastic response of the soft component.
In our calculations, both hard and soft rods have a hard core of equal diameter.

(ii) For a given value of $d$, the phase gap at $IN$ coexistence
is significantly smaller for systems with smaller values
of $k$ than that for large $k$. A large phase gap is usually
seen as indicative of polydispersity effects. Interestingly, a
narrow $IN$ phase gap was observed in the experiments with PEG-coated
fd-viruses \cite{PVGJFPRL05}, much narrower than to be expected
if both components were indeed  incompressible.

(iii) Although a rigidity of $k=10$ seems high and should render
the mixture close to that of hard rods of unequal diameter,
as can inferred from our results on monodisperse rods in Sec.~\ref{monosect},
a $N_{1}N_{2}$ phase separation is not observed for $d=4.0$. We recall that
it does manifest itself in rigid hard-rod mixtures of this diameter
ratio \cite{RMDP98,SRPRE03}. Such a strong sensitivity of
the stability of the nematic phase on the elasticity of the shell is also seen in Fig.~\ref{bin}(c),
where an increase of the rigidity from $k=5$ to
$k=10$ hardly affects the $IN$ transition curves whereas it does prevent the
$N_{1}N_{2}$ remixing at high pressures in the $k=10$ case.

Finally, in Fig.~\ref{gam43} we show the compression factor $\gamma_{I,N}(x)$ of the polymeric
shells of the rods in the coexisting $I$ and $N$ phases for
$k=0.1,1,5,10$ for $d=4.3$, as a function of the mole fraction $x$.
Mixtures with other values of $d$ and $k $ produce
similar curves. Upon an increase of the relative concentration $x$ of the soft
rods, compression of the polymer shells increases monotonically, which
reflects proportionality of the elastic energy to $x$. Note that
$\gamma_{I}(x) <\gamma_{N}(x)$ for all the systems studied.
Almost linear curves $\gamma_{I,N}(x)$ for extremely soft rods
($k=0.1$) again indicate effectively monodisperse behavior.

\section{Summary and discussion}\label{nasummarysect}
In this paper we have explored the bulk phase diagrams of
monodisperse compressible rods and those of binary mixtures of
incompressible and compressible rods. Our motivation for this are
the experiments by Purdy and coworkers, in which polymer-coated fd
virus particles were mixed with bare fd in aqueous suspension
\cite{PVGJFPRL05}. Not surprisingly, our study shows that a pure
system of coated rods, modeled here by an elastically responding
excluded-volume interaction, requires a higher number density to obtain a
nematic phase than that of incompressible rods of equal diameter.
Nevertheless, the density ratio of the rods in the coexisting
phases remains very close to that of the pure hard-rod system,
being $1.274$. The same, in fact, is true for the nematic order
parameter. This implies that these dimensionless quantities cannot
be used to estimate the elastic modulus $k$ of the polymer
coating. For known values for the coat thickness, $\Delta^{pol}$,
and the bare core diameter, $\Delta^{core}$, that of the modulus
$k$ might be calculated from the observed absolute density in
either coexisting phase, using, e.g., Fig.~\ref{phaseD3} if
$\Delta^{pol}/\Delta^{core} = 2$.

The situation is very different when it comes to mixtures of
compressible and incompressible rods, which we have investigated
for relative coat thicknesses in the range of values equal to
$d=1+ \Delta^{pol}/\Delta^{core} =3.0$, $4.0$, $4.3$ and $4.5$. We
find that the topology of the phase diagram in the
pressure-composition representation changes quite dramatically in
this small range of $d$-values, with $0.1 \leq k \leq 10$. Upon an
increase of the shell rigidity from the lowest to the highest
value, we find (i) that the degree of fractionation at $IN$
coexistence increases, and (ii) that the $NN$ binodal, if present,
changes from a dome like to chimney like, i.e., the upper critical
point moves to infinite pressures at a critical value of $k$ that
depends on $d$. For conditions where there is an upper critical
point in the phase diagram, we did not find a reentrant $NN$ phase
separation at higher pressures, i.e., no lower critical $NN$ point
was found. Therefore, the experimentally observed lower critical
point cannot be explained by a compressible polymer coating, at
least not within an Onsager type of approach.

As it is now clear that the phase behavior of the rods depends
sensitively on the presence and properties of a polymeric coating,
we suggest that a realistic model of any experimental system
involving coated rod-like particles should not only correct for a
finite bending rigidity and/or finite-length, but also for the
elastic softness of that coating. This we intend to pursue in the
near future.

\begin{acknowledgments}
This work is part of the research program of the `Stichting voor
Fundamenteel Onderzoek der Materie (FOM)', which is financially
supported by the `Nederlandse organisatie voor Wetenschappelijk
Onderzoek (NWO)'.
\end{acknowledgments}


\begin{thebibliography}{99}
\bibitem {comIN}
It was proven in \cite{O} for the isotropic phase, and
generally accepted for the nematic phase. See, e.g. \cite{VLRPP92}.
\end{thebibliography}
\end{document}